\documentclass[amssymb,11pt]{article}
\usepackage[pdftex,colorlinks=true,urlcolor=black,filecolor=black,linkcolor=black,
            pdftitle={New Five Dimensional Spherical Vacuum Solutions.},
            pdfauthor={Mark D. Roberts},
            pdfsubject={Five Dimensions, Vacuum Solutions, Kaluza-Klein},
            pdfkeywords={five dimensions,signature,vacuum solutions},
            pagebackref,pdfpagemode=None,bookmarksopen=true]{hyperref}
\usepackage{amssymb}
\newcommand{\bc}{\begin{center}}
\newcommand{\ec}{\end{center}}
\newcommand{\be}{\begin{equation}}
\newcommand{\ee}{\end{equation}}
\newcommand{\ber}{\begin{eqnarray}}
\newcommand{\ear}{\end{eqnarray}}

\newcommand{\bx}{\Box}

\newcommand{\Lg}{{\cal L}}
\newcommand{\n}{\nonumber\\}

\newcommand{\sig}{{\bf s}}

\begin{document}
\title{New Five Dimensional Spherical Vacuum Solutions.}
\author{Mark D. Roberts,\\
\href{http://www.ihes.fr}
     {Institut Des Hautes {\'E}tudes Scientifiques},
le Bois-Marie,  35,  Route de Chartres,\\
Bures-sur-Yvette,
France,
F-91440.\\
mdr@ihes.fr
}
\maketitle
\vspace{1.3in}
\begin{abstract}
A new five dimensional spherical vacuum solution is both dervied and
its signature,  curvature and truncation discussed.
Its truncation leads to a four dimensional spacetime with similiar stress to those found
by charge-free Kaluza-Klein compactification.
Various other restrictions to four dimensions are looked at to see if they have stresses
consisting of electromagnetic fields or quadratic tensors.
The solution is extended to the brane picture where the extended space is found to obey
field equations with metric stress derivable from a lagrangian dependent on brane function
and kaluza scalar.
\end{abstract}
\newpage
\tableofcontents
\newpage
\section{Introduction.}\label{intro}

\subsection{Motivation.}\label{motivation}
There seems to be two main types of motivation for studying higher dimensional exact solutions.
The first is the {\em astrophysical} motivation in which higher dimensions produce a small
modification of results already known in four dimensions.  An example of this is that the
Jebsen-Birkhoff theorem \cite{jebsen,birkhoff,DF,JR} no longer holds in higher dimensions
\cite{BS,BM,KG} allowing modifications of schwarzschild spacetime which can be tested against
standard observations.  Another example is that there might be modifications to cosmology
\cite{mdrgb,mdrkb}.
The second is the {\em process} motivation in which higher dimensions produce new processes
which do not occur in four dimensions.  A possible example of this could be the transfer of
a scalar field between four dimensional spacetime and higher dimensions \cite{mdrfi},  or
the transfer of gravitons \cite{LS}.

\subsection{Examples.}\label{examples}

There are examples of generalizations of the schwarzschild solution in higher dimensions
\cite{tangherlini,CD} and of how it an be embedded in brane theory \cite{BCG}.
The homogeneous schwarzschild solution is
\be
ds_5^2=-\left(1-\frac{2m}{r^2}\right)dt^2+\left(1-\frac{2m}{r^2}\right)^{-1}dr^2+r^2d\Sigma_3^2,
\label{hsch}
\ee
where
\be
d\Sigma_3^2\equiv d\chi^2+\sin(\chi)^2d\Sigma_2^2,~~~
d\Sigma_2^2\equiv d\theta^2+\sin{\theta}^2d\phi^2,
\label{ang}
\ee
note the power of $r$ in the first two terms of ref{hsch}.
The minimally extended schwarschild solution is
\be
ds_5^2=-\left(1-\frac{2m}{r}\right)dt^2+\left(1-\frac{2m}{r}\right)^{-1}dr^2+r^2d\Sigma_2^2+d\chi^2,
\label{msch}
\ee
however
\be
ds_5^2=-dt^2+\left(1-\frac{2m}{r}\right)^{-1}dr^2+r^2d\Sigma_2^2+\left(1-\frac{2m}{r}\right)d\chi^2,
\label{rsch}
\ee
is also a solution.
The Vaidya form of \ref{hsch} has Kretschmann curvature invariant
$K\equiv R_{abcd}R^{abcd}=288m^2/r^8$,
while the Vaidya form of \ref{msch}
has the same value as the four dimensional case $K=48m^2/r^6$.
The sort of solution that one would hope for is like \ref{msch}
but with a small fifth term of the form of \ref{rsch},
as this might allow comparison with astrophysical observations.
Such a solution has not been found yet,
however a solution midway between \ref{msch} and \ref{rsch} has line element
\be
ds_5^2=\left(1-\frac{2m}{r^2}\right)^{-1}dr^2+r^2d\Sigma_2^2
-\frac{1+\sqrt{1-\frac{r^2}{2m}}}
      {1-\sqrt{1-\frac{r^2}{2m}}}dt^2
+\frac{1-\sqrt{1-\frac{r^2}{2m}}}
      {1+\sqrt{1-\frac{r^2}{2m}}}d\chi^2.
\label{midway}
\ee

\subsection{Conventions \& Outline.}\label{conv}

The conventions used are those of grtensor \cite{MPL}.
Truncation means when higher dimensional components are ignored,
in present circumstances compactification means a procedure which associates higher dimensional
metric coefficients with lower dimensional fields;
$y$-truncation,  $y$ compactification means when those procedures
are carried out with $g_{55}=1$;  similarly $\chi$-truncation is when $g_{55}\ne$ a constant,
then typically $g_{55}$ is a scalar field.
Constants $A,B,\ldots,\alpha,\beta,\ldots,k,l,\ldots$ are used repeatedly and hopefully no
ambiguity occurs.  Metric indices $a,b,\ldots$ are the same for both four and five dimensions,
which should be clear from the context.
The derivation of the midway solution \ref{midway} is discussed in \S\ref{der},
its properties are discussed in \S\ref{prop},
and its truncation to four dimensional spacetime is discussed in \S\ref{trun}.
\S\ref{fdr} discusses off-diagonal five dimensional spaces and their truncation.
\S\ref{braneex} discusses extension of the solution \ref{midway} to the brane picture.
\S\ref{conc} is the conclusion.

\section{Derivation.}\label{der}

The five dimensional static spherical line element is taken to be
\be
ds_5^2=-\exp(2\nu(r))dt^2+\exp(2\lambda(r))dr^2+r^2d\Sigma_2^2+\exp(2\psi(r))d\chi^2.
\label{fived}
\ee
The non-vanishing components of the Ricci tensor are given by
\ber
-R_{rr}&=&-\frac{2\lambda'}{r}+\nu"-\lambda'\nu'+\nu'^2+\psi"-\lambda'\psi'+\psi'^2,\n
R_{\theta\theta}&=&1+\exp(-2\lambda)\left(-1+r(\lambda'-\nu'-\psi')\right)\n
\exp(2(\lambda-\nu))R_{tt}&=&\nu"-\lambda'\nu'+\nu'^2+\frac{2\nu'}{r}+\nu'\psi'\n
-\exp(2(\lambda-\psi))R_{\chi\chi}&=&\psi"-\lambda'\psi'+\psi'^2+\frac{2\psi'}{r}+\nu'\psi'.
\label{afe}
\ear
Defining
\be
f\equiv\nu',~~~
g\equiv\lambda',~~~
h\equiv\psi',
\label{deffgh}
\ee
gives
\ber
-R_{rr}&=&-\frac{2g}{r}+f'-gf+f^2+h'-gh+h^2\n
R_{\theta\theta}&=&1+\exp(-2\lambda)\left(-1+r(g-f-h)\right),\n
\exp(2(\lambda-\nu))R_{tt}&=&f'-gf+f^2+\frac{2f}{r}+fh\n
-\exp(2(\lambda-\psi))R_{\chi\chi}&=&h'-gh+h^2+\frac{2h}{r}+fh
\label{ricfgh}
\ear
forming
\be
-R_{rr}-\exp(2(\lambda-\nu))R_{tt}+\exp(2(\lambda-\psi))R_{\chi\chi}=-\frac{2}{r}(g+f+h)-2fh
\label{gish}
\ee
and assuming
\be
R_{ab}=lg_{ab}
\label{fel}
\ee
\ref{gish} gives an expression for $g$
\be
g=-f-h-rfh-\frac{1}{2}lr\exp(2\lambda),
\label{gisf}
\ee
so that it can now be eliminated;
substituting back into \ref{ricfgh} and assuming \ref{fel}
\be
0=f'+2f^2+\frac{2f}{r}+(hf+\frac{1}{2}l\exp(2\lambda))(2+rf),
\label{fh}
\ee
and an identical equation with $f$ and $h$ interchanged.
Note that the last term can be thought of as an interaction term.
For simplicity choose
\be
h=kf,~~~l=0,
\label{kdef}
\ee
substituting \ref{kdef} into \ref{fh} gives one equation
\be
0=f'+\frac{2f}{r}+2(1+k)f^2+krf^3.
\label{feq}
\ee
Maple gives solution to this which contains $f$ inside expressions
which are not invertible for $f$ and so $f$ cannot be integrated to give a metric function.
The case that $k=0$ gives \ref{msch} and $1/k=0$ gives \ref{rsch},
the only other case so far found with invertible $f$ is for $k=-1$,  then
\be
f=\frac{\pm1}{r\sqrt{1+Cr^2}},
\label{ffound}
\ee
integrating
\be
\nu=\int f dr=-{\rm arctanh}\frac{1}{\sqrt{1+Cr^2}},
\label{nufound}
\ee
converting arctanh to logarithmic form and taking the constant $C=-2m$ gives
the midway solution \ref{midway} in original coordinates,
although it is perhaps better referred to as the
reciprocal midway solution as $g_{\chi\chi}=-1/g_{tt}$.

\section{Properties.}\label{prop}

\subsection{Signature}\label{sig}

First consider what happens to schwarzschild geometry when signature constants,
$\{\sig_i=+1~{\rm or}-1\}$,  are added
\be
ds^2=\frac{\sig_rdr^2}{\left(1-\frac{2m}{r}\right)}+\sig_\theta r^2d\theta^2
+\sig_\phi r^2\sin(\theta)^2d\phi^2+\sig_t\left(1-\frac{2m}{r}\right)dt^2,
\label{sigsch}
\ee
the Ricci tensor now has non-vanishing components
\be
R_{\theta\theta}=\left(1-\frac{\sig_\theta}{\sig_r}\right),~~~
R_{\phi\phi}=\sin(\theta)^2\left(1-\frac{\sig_\theta}{\sig_r}\right)\frac{\sig_\phi}{\sig_\theta},
\label{sigsr}
\ee
so that the solution remains a vacuum solution if $\sig_r=\sig_\theta$ while $\sig_\phi,\sig_t$
remain free.   A choice of sign of $\sig_r$ is equivalent to a choice of convention for $ds^2$,
so that there are four independent choices left from $\{\sig_\phi=(+1,-1),\sig_t=(+1,-1)\}$.
The Kretschmann curvature invariant remains unchanged $K=48m^2/r^6$ for all four choices.

Now consider what happens to the midway solution \ref{midway} when signature constants are added
\ber
ds_5^2&=&\sig_r\left(1-\frac{2m}{r^2}\right)^{-1}dr^2
+\sig_\theta r^2d\theta^2+\sig_\phi r^2\sin(\theta)^2d\phi^2\n
&&+\sig_t\frac{1+\sqrt{1-\frac{r^2}{2m}}}
            {1-\sqrt{1-\frac{r^2}{2m}}}dt^2
+\sig_\chi\frac{1-\sqrt{1-\frac{r^2}{2m}}}
               {1+\sqrt{1-\frac{r^2}{2m}}}d\chi^2,
\label{sigmidway}
\ear
the non-vanishing components of the Ricci tensor are again given by \ref{sigsr},
leaving eight independent choices of signature
$\{\sig_\phi=(+1,-1),\sig_t=(+1,-1),\sig_\chi=(+1,-1)\}$.
The choice of $\sig_\phi$ will not be considered anymore here,  leaving four choices.
Note in particular the original choice $-++-+$ and the choice $-++++$ are allowable.
If one wants to truncate to a spherical spacetime metric with $-+++$ then there are
two ways left to do this,  corresponding to either
$({1+\sqrt{1-\frac{r^2}{2m}}})/({1-\sqrt{1-\frac{r^2}{2m}}})$
or
$({1-\sqrt{1-\frac{r^2}{2m}}})/({1+\sqrt{1-\frac{r^2}{2m}}})$
in a metric function;  the resulting spacetimes turn out to be equivalent
by the coordinate transformation \ref{cshift},
therefore the five dimensional solution \ref{midway} has a unique truncatation
four dimensional spacetime.

\subsection{Curvature}\label{curv}

In the coordinate system \ref{midway} the Kretschmann curvature invariant is
\be
K=\frac{96m(3m-r^2)}{r^8},
\label{midK}
\ee
the Gauss-Bonnet tensor takes a simple form:
the Ricci,  Einstein,  Bach and other quadratic tensors vanish.

\subsection{Killing vector.}\label{kv}

In addition to the killing coordinates $t,\phi,\chi$
there is the expansion free killing vector
\be
KV_{a}=\frac{A}{r^2}\left(1+\sqrt{1-\frac{r^2}{2m}}\right)^2\delta_a^t
               +Br^2\left(1+\sqrt{1-\frac{r^2}{2m}}\right)^{-2}\delta_a^\chi,
\label{killing}
\ee
where $A$ and $B$ are constants.

\subsection{Alternative forms of the metric.}\label{alt}

The midway solution \ref{midway} contains a square root,
in order to investigate what happens when $r^2>2m$
one can look at alternative forms of the metric;
if $\sqrt{1-r^2/(2m)}\rightarrow \sqrt{r^2/(2m)-1}$
then the metric is no longer a vacuum solution.
For $m>0$ defining
\be
r^2\equiv 2m-r'^2,
\label{mshift}
\ee
and dropping the prime,  the metric takes the shifted form
\be
ds_5^2=-dr^2+(2m-r^2)d\Sigma_2^2
-\frac{\sqrt{2m}+r}{\sqrt{2m}-r}dt^2+\frac{\sqrt{2m}-r}{\sqrt{2m}+r}d\chi^2,
\label{midshift}
\ee
for $m<0$ define
\be
r^2\equiv2m+r'^2,
\label{mmshift}
\ee
again dropping the prime
\be
ds_5^2=+dr^2+(2m+r^2)d\Sigma_2^2
-\frac{\sqrt{-2m}+r}{\sqrt{-2m}-r}dt^2+\frac{\sqrt{-2m}-r}{\sqrt{-2m}+r}d\chi^2.
\label{mmidshift}
\ee
Using the shifted firm radial coordinate to define the trigonometrical radial coordinate
\be
r^2\equiv2m\cos(\alpha)^2,
\label{mtrig}
\ee
the metric takes trigonometric form
\be
ds_5^2=2m\sin(\alpha)^2\left(-d\alpha^2+d\Sigma_2^2\right)
-\cot(\alpha/2)^2dt^2+\tan(\alpha/2)^2d\chi^2,
\label{midtrig}
\ee
where the trigonometrical relationship
\be
\cos(t)=1-2\sin(t/2)^2,
\label{teeovertwo}
\ee
is used to simplify the last two terms.
In this form the coordinate transformation
\be
\cos(\alpha)=-\cos(x)
\label{cshift}
\ee
flips the sign pattern in the last two terms,
providing the coordinate transformation that shows that
the truncation to four dimensional spacetime is unique,
subject to the requirements:  the line element is diagonal,  the signature is correct and
that there is no dependence in the metric coefficients on the fifth dimensional coordinate.
There is a hyperbolic form
\be
ds_5^2=
-\coth(r/2)^2dt^2
+2m\sinh(r)^2(dr^2+d\Sigma_2^2)
+\tanh(r/2)^2d\chi^2,
\label{midhyper}
\ee
which differs from the trigonometric form \ref{midtrig} in the relative signs of the
$dr^2$ and $d\Sigma_2^2$ terms.
Using combined notation ${\rm s}(t)=\sin(t)~or\sinh(t)~etc.$
the combined trigonometric \ref{midtrig} and hyperbolic \ref{midhyper} form of \ref{midway} is
\be
ds_5^2=A{\rm ct}(t/2)^2dr^2+2Bm{\rm s}(t)^2d\Sigma_2^2+2Cm{\rm s}(t)^2dt^2
+D{\rm tn}(t/2)^2d\chi^2,
\label{combmid}
\ee
which is a vacuum solution when
\be
0=B+\epsilon C,~~~
\epsilon=\left\{+1 {\rm ~for~trig},-1 {\rm ~for~hyper}\right\}.
\label{combmidsig}
\ee
The hyperbolic form leads to the $y$ form
\be
ds_5^2=+\frac{\sqrt{2m}+y}{\sqrt{2m}-y}dt^2-\frac{\sqrt{2m}-y}{\sqrt{2m}+y}dr^2
+(y^2-2m)d\Sigma_2^2+dy^2,
\label{midwayy}
\ee
in which one can think of $y$ rather than $\chi$ as the fifth coordinate;
however up to signature this line element is the same as the shifted form \ref{mshift}.
None of the alternative forms of the metric seem to extend the solution \ref{midway} to $r^2>2m$.

\subsection{Limits}\label{lim}

Consider the metric in the form \ref{midway}:
in the limit $r^2\rightarrow0$ it goes to $\{-\infty,r^2,r^2\sin(\theta)^2,-1,+1\}$;
the limit $r^2\rightarrow 2m$ gives $K=6/m^2$,
so it is not clear what could happen for $r^2>2m$,
the limit $2m\rightarrow0$ depends on which coordinate system is used;
in \ref{midway} it is not clear what happens as $r$ must remain smaller than $2m$,
in \ref{midshift} again it is not clear what happens if $r$ becomes smaller than $2m$
however if this is relaxed then the limit gives$\{-1,-r^2,-r^2\sin(\theta)^2,+1,-1\}$,
in \ref{midtrig} the three first terms just disappear, allowing speculation that these
could be used in a model of dimensional creation and annihilation.

\section{Truncation to four dimensional spacetime.}\label{trun}

\subsection{$y$-truncation}\label{midytrun}

First consider $y$-truncation of the line element \ref{midwayy}:
the y-truncated metric is the same as \ref{midwayy} but without the final $dy^2$,
it has curvature
\be
K_4=\frac{4}{(y^2-2m)^2},~~~
K_5=6mK_4^2,
\label{yK}
\ee
however it has four dimensional metric components dependent on the fifth dimensional coordinate,
and no other coordinate, so it is not considered here anymore.

\subsection{$\chi$-truncation}\label{midchitrun}

$\chi$-truncation to four dimensions in which there is no dependence on the fifth dimensional coordinate,
and the four dimensional spacetime has the correct signature is unique.
Truncating the metric in the form \ref{midtrig} with
$\{\alpha,\theta,\phi,t,\chi\}\rightarrow\{t,\theta,\phi,ir,.\}$ gives the line element
\be
ds_4^2=2m\sin(t)^2\left(-dt^2+d\Sigma_2^2\right)+\tan(t/2)^2dr^2.
\label{trunmid2}
\ee
The metric has killing coordinates $r,\phi$ and has the killing vector
\be
KV_{a}=A\tan(t/2)^2\delta_a^r,
\label{trunkv}
\ee
which has non-vanishing acceleration,  shear, electric part of the weyl tensor,
and vanishing expansion scalar,  vorticity vector,  and magnetic part of the weyl tensor.
The curvature of this line element is described by
\be
WeylSq=\frac{3}{4m^2\sin(t)^4\cos(t/2)^4},
\label{weylsqtchi}
\ee
and the Ricci tensor
\be
R_{a.}^{~b}=\frac{1}{4m\sin(t)^4}{\rm diag}\left[1,\cos(t),\cos(t),-1-2\cos(t)\right],
\label{trunricci}
\ee
which is tracefree,  $K$ can be constructed from \ref{weylsqtchi} and \ref{trunricci},
however it is not a function of $K_5$ as part of the $t$-dependence is in $g_{55}$ and
this is lost in truncation.   Quadratic tensors take a simple form but so far have not led
to any field equations.   The potential
\be
V=\cot(t/2)
\label{trunpot}
\ee
can be used to express the Ricci tensor as
\be
R_{ab}=\frac{V_{(a;b)}}{V}.
\label{trunscalar}
\ee
The vector $V_a$ has non-vanishing acceleration,  shear,  electric part of the weyl tensor
and vanishing vorticity,  expansion scalar and magnetic part of the weyl tensor.
The metric stress \ref{trunscalar} can be derived from the lagrangian
\be
\Lg=\frac{\bx V}{2V},
\label{trunslag}
\ee
this can be thought of as the scalar field from Kaluza \cite{kaluza} compactification,
which is $V^2=g_{55}$.
There is no electromagnetic field as there are no off-diagonal $g_{5i}$ terms
in the five dimensional space.
Adding enough fields any four dimensional spherical spacetime obeys
field equations \cite{mdr13}Appendix I, last equation.

\section{Five dimensions revisited.}\label{fdr}

\subsection{Radial off-diagonal solution.}\label{offdr}

By trial and error the vacuum radial off-diagonal line element
\ber
ds_5^2&=&-\tan(r/2)^2dt^2+2m\sin(r)^2\left(dr^2+d\Sigma^2_2\right)\n
&&+\cot(r/2)^2d\chi^2+8\sqrt{m}\cos(r/2)^2drd\chi,
\label{offdiag}
\ear
was found.   It is similar to \ref{midtrig} except for the interchange of $r$ and $t$ and the
off-diagonal term.
It is also expressible in terms of the luminosity coordinate $r'^2\equiv 2m\sin(r)^2$
and various coordinate systems corresponding to those of the previous solution,
however for calculational purposes the easiest form of the metric to work with is \ref{offdiag}
with the $r/2$ trigonometric functions replaced by their $r$ form using \ref{teeovertwo}.
Alternative forms of the metric are discussed in \S\ref{gencase} below.

In addition to the killing coordinates $t,\phi,\chi$
there is the expansion free killing vector
\be
KV_{a}=\frac{A\sin(r/2)^4}{\sin(r)^2}\delta_a^t,
\label{oddkilling}
\ee
where $A$ is a constant.

\subsection{Radial off-diagonal $\chi$-truncation.}\label{odctrun}

$\chi$-truncation can be implemented by discarding the fifth dimensional terms
of the line element \ref{offdiag}.   The Ricci tensor and scalar are
\ber
R_{ab}=\frac{1}{\sin(r)^2}{\rm diag}
\left[\alpha,\beta,\sin(\theta)^2\beta,\gamma\right],&&
R=\frac{2}{m\sin(r)^2},\\
\alpha\equiv1+2\cos(r),~~~
\beta\equiv2\sin(r)-\cos(r),&&
\gamma\equiv\left(2m(2+2\cos(r)-\sin(r)^2)\right)^{-1}.
\nonumber
\label{odricci}
\ear
If identification with an electromagnetic stress is attempted then kaluza compactification
does not work:  identifying $g_{r5}$ proportional to $A_r$ one finds that $A_r$ does not
explicitly contribute to the stress,  only $A_t$ does.
The quadratic tensors take simple form but appear not to obey any simple field equations.
The $\chi$-truncation of \ref{offdiag} produces a line element of the same form
as \ref{trunmid2} but with different signature,  this shows up if for example
field equations such as \ref{trunscalar} are sought in that the $\theta\theta$ component
becomes non-vanishing,  non-vanishing $\theta\theta$ component is usually and indication that
the relative signs of $g_{rr}$ and $g_{\theta\theta}$ are wrong.  So now the case of almost
general signature is looked at,  'almost' in the sense that the relative signs of
$g_{\theta\theta}$ and $g_{\phi\phi}$ are fixed and everything else is free.

\subsection{General case.}\label{gencase}

Generalizing the line element in the shifted form \ref{midshift} gives
\be
ds_5^2=A\frac{k+t}{k-t}dr^2+B(k^2-t^2)d\Sigma_2^2+Cdt^2
+D\frac{k-t}{k+t}d\chi^2+\sqrt{E\frac{t-k}{t+k}}dtd\chi,
\label{genle}
\ee
where $k=\sqrt{2m}$ for $m$ positive and $k=\sqrt{-2m}$ for $m$ negative,
this is a vacuum solution when
\be
(B+C)D+E=0
\label{genvs}
\ee
choosing the solution
\be
C=-B-\frac{E}{D}
\label{constraint}
\ee
the curvature invariant is found to be
\be
K=\frac{24k^2(2t^2+k^2)}{B^2(t^2-k^2)^4},
\label{Kgen}
\ee
as this does not contain $E$ it suggests that the diagonal and off-diagonal solutions
are the same,  and this turns out to be the case,  performing the coordinate transformation
\be
\chi\rightarrow\chi'+\frac{\sqrt{E}}{D}I,~~~
I\equiv\int\sqrt{\frac{k+t}{k-t}}dt
=-\sqrt{k^2-t^2}+k\arctan\frac{t}{\sqrt{k^2-t^2}},
\label{tcoordtrans}
\ee
on the diagonal line element \ref{mshift} produces \ref{genle}.
Truncating \ref{genle} the field equations \ref{trunscalar} are obeyed when $C+B=0$.
It is possible to show that there are no field equations involving non-zero electromagnetic
field and scalar fields which have line element the four dimensional truncation of \ref{genle}.

Performing linear coordinate transformations in $\chi$ and $r$ gives the two line elements
\be
ds_5^2=\frac{2Af(t)}{k^2-t^2}dr^2+B(k^2-t^2)d\Sigma_2^2+Cdt^2
+2\sqrt{E}\frac{k-t}{k+t}drd\chi+D\frac{k-t}{k+t}d\chi^2.
\label{roffdiag}
\ee
For the first
\be
f_1(t)\equiv(t^2+k^2),~~~E-AD=0,~~~C+B=0,
\label{first}
\ee
when $1=A=B=-C=D$ the truncated Ricci tensor is
\be
R_{a.}^{~b}=\frac{2k^2}{(k^2-t^2)^2(t^2+k^2)^2}{\rm diag}\left[k^4+t^4,t^2,t^2,2t^4\right],~~~
R=\frac{2k}{(t^2+k^2)^2}.
\label{firstricci}
\ee
For the second
\be
f_2(t)\equiv2kt,~~~E+AD=0,~~~C+B=0,
\label{second}
\ee
$E$ occurs inside a square root,  real metric coefficients then imply that either $A$ or $D$
is negative, choosing $D$ negative implies signature $+,+,+,-,-$,
explicitly $1=A=B=-C=-D$ then the truncated Ricci tensor is
\ber
R_{a.}^{~b}=\frac{1}{4t^2(k^2-t^2)^2}{\rm diag}\left[\alpha,\beta,\beta,\gamma\right],~~~
&&R=-\frac{1}{2t^2},\\
\alpha\equiv(t^2-2kt-k^2)(-t^2-2kt+k^2),~~~
\beta\equiv-2t^2(t^2+k^2),
&&\gamma\equiv(3t^2-k^2)(t^2+k^2).
\nonumber
\label{secondricci}
\ear
So far performing linear coordinate transformations on $\chi,\theta,\phi$
has not produced anything as simple as the above.

\section{Brane extension.}\label{braneex}

Introduce the five dimensional metric
\be
ds_5^2=U(\chi)^2ds_4^2+V(x^4)^2d\chi^2,
\label{branele}
\ee
where $U$ is the brane function and $V$ is the kaluza scalar.
Choose $U=\exp(\chi/\alpha)$,  $V$ given by \ref{trunpot},  $ds_4^2$ given by \ref{trunmid2},
then the metric obeys the field equations
\be
R_{ab}=-\frac{4V^2}{\alpha^2}g_{ab}-\frac{6}{UV}U_{(a}V_{b)},
\label{branericci}
\ee
this matter stress can be derived from the lagrangian
\be
\Lg=-\frac{6V^2}{\alpha^2}-\frac{3}{UV}U_{(a}V^{a)}.
\label{branelag}
\ee

\section{Conclusion.}\label{conc}

\subsection{Signature.}{\label{concsig}

The first thing to note about the solution in its original form \ref{midway} is its signature:
it is $-,-,+,+,+$.  Spaces with two times are usually considered unphysical as even the flat
case allows closed time-like curves.  To discuss possible signatures the method of
signature constants was introduced \ref{sig},  this entailed setting all metric components
positive and multiplying each by a constant,  here called a signature constant.  Then the Ricci
curvature is recomputed to produce an algebraic equation which combines signature constants
and has to hold if the space is to remain a vacuum solution,
see for example \ref{sigsr},\ref{combmidsig},\ref{genvs}.
There often turns out to be several choices,  however the requirements of $\chi$-truncation
to a four dimensional spacetime with physical signature entails a unique choice.
Some choices of signature result in spaces in which the angular terms are not multiplied by
a spacelike luminosity radial coordinate,  all that has happened is the the Takeno type
\cite{takeno} has changed from type I for which there is angular term $r^2d\Sigma_2^2$
to Takeno type II for which there is the angular term $r^0d\Sigma_2^2$.   An example of Takeno
type II spacetimes are the Bertotti-Robinson solutions \cite{bertotti,robinson,DL}.
There is the question of how many new solutions there actually are:  \ref{genle} and
\ref{genvs} cover all cases so in that sense there is only one,   on the other hand one could
think of each set of signature constants representing a separate solution.
There seems to be no direct application of the method of signature constants to the theory
of the line element field \cite{mdr36}, but with further study this may change.

\subsection{Curvature.}\label{conccurv}

The decay of the Kretschmann curvature invariant $K$ was proportional to $r^{-2(d-1)}$,
see \ref{midK},
as would be expected by analogy with the homogeneous schwarzschild solution \ref{hsch}.
y-truncation \ref{midytrun} produced an unanticipated relationship \ref{yK} between the
four and five dimensional Kretschmann curvature invariant.

\subsection{Field equations.}\label{conctc}

Having a five dimensional solution is only part of the problem:
also one needs a four dimensional surface in it to model spacetime.
The choice is simplified if there is a requirement that the four dimensional
surface's metric coefficients have no dependence on the fifth coordinate:
this is sometimes called the cylinder condition.
Subject to this condition,  and that the metric is diagonal,
and also that four dimensional spacetime has physical signature,
the truncation of \ref{midway} is unique and has stress given by a scalar field
\ref{trunpot},\ref{trunscalar} similar to that in Kaluza-Klein theory.
Attempts to use a different four dimensional surface corresponding to
a spacetime with an electromagnetic field,
or that obey higher order field equations have so far been unsuccessful.

\subsection{Brane picture.}\label{branepic}

The exact spherical five dimensional vacuum solution in the form \ref{midtrig},\ref{trunmid2}
can have four dimensional components multiplied by a brane function to produce a line element
of the form \ref{branele}.   This line element obeys field equations with metric stress
\ref{branericci} and lagrangian \ref{branelag};  at the moment this appears to be just
coincidence,  there seems to be no a priori reason why this should be so.

\subsection{Objectives.}\label{concobj}

The astrophysical motivation expressed in the introduction \ref{motivation}
is not met as the new solutions presented here \ref{genle},\ref{branele}
are not similar to familiar four dimensional solutions.
Technically the reason for this is the choice \ref{kdef} with $k=-1$,
for the solution to be near the schwarzschild solution $k$ would have to be small;
there is the possibility of producing small $k$ perturbation theory,
but this would lead us away from the realm of exact solutions.
The process motivation expressed in the introduction \ref{motivation} is not directly met as
no new physical process is predicted,  however the vacuum solutions \ref{midway} and
\ref{genle} exhibit the properties described above which illustrate five dimensional
behaviour which could hint at new processes.


\end{document}